\documentstyle[12pt,epsfig]{article}

\newcommand{\be}{\begin{equation}}
\newcommand{\ee}{\end{equation}}
\def\aprle{\buildrel < \over {_{\sim}}}
\def\aprge{\buildrel > \over {_{\sim}}}
\begin{document}
\topmargin 0pt
\oddsidemargin=-0.4truecm
\evensidemargin=-0.4truecm
\renewcommand{\thefootnote}{\fnsymbol{footnote}}
\newpage
\setcounter{page}{1}
\begin{titlepage}     
\vspace*{-2.0cm}
\begin{flushright}
FISIST/1-99/CFIF \\
hep-ph/9903302
\end{flushright}
\vspace*{0.5cm}
\begin{center}
{\Large \bf Remarks on parametric resonance of neutrino \\ 
\vspace*{0.2cm}
oscillations in the earth}
\vspace{1.0cm}

{\large E. Kh. Akhmedov
\footnote{On leave from National Research Centre Kurchatov Institute, 
Moscow 123182, Russia. E-mail: akhmedov@gtae2.ist.utl.pt}}\\
\vspace{0.05cm}
{\em Centro de F\'\i sica das Interac\c c\~oes Fundamentais (CFIF)} \\
{\em Instituto Superior T\'ecnico}\\
{\em Av. Rovisco Pais, P-1096 Lisboa-Codex, Portugal }\\
\end{center}
\vglue 0.8truecm
\begin{abstract}
Neutrino oscillations in matter can exhibit a specific 
resonance enhancement -- parametric resonance, which is different 
from the MSW resonance. Recently it has been shown that the 
oscillations of atmospheric and solar neutrinos inside the earth 
can undergo parametric enhancement when neutrino trajectories cross 
the core of the earth. In this paper we continue 
the study of the parametric resonance of neutrino oscillations 
in the earth. The following issues are discussed: 
stability of the resonance with respect to the variations of the zenith angle 
of the neutrino source; higher-order resonances; prospects of the experimental 
observation of the parametric resonance of neutrino oscillations.  
We also comment on a recent controversy regarding the physical nature 
of the resonance enhancement of the oscillations of the core crossing 
neutrinos in the earth.   
\end{abstract}
\end{titlepage}
\renewcommand{\thefootnote}{\arabic{footnote}}
\setcounter{footnote}{0}
\newpage
\section{Introduction}
It is well known that neutrino oscillations in matter can differ significantly  
from oscillations in vacuum, the best studied 
example being the 
Mikheyev-Smirnov-Wolfenstein (MSW) effect \cite{MS,W}. 
The MSW effect is an enhancement of the neutrino flavor transition 
probability due to an amplification of the oscillation 
{\em amplitude} in matter. However, possible enhancement effects of matter on 
neutrino oscillations are not limited to the MSW effect. It has been 
pointed out in \cite{ETC,Akh1} that the probability of neutrino flavor 
transition can also be strongly enhanced if the oscillation {\em phase} 
undergoes certain modification in matter. This can happen if the change of the 
matter density along the neutrino path is correlated in a certain way with the 
change of the oscillation phase. This amplification of the neutrino oscillation 
probability in matter due to specific phase relationships has an interesting 
property that it can accumulate if the matter density profile along the neutrino 
path repeats itself, {\it i.e.} is periodic. The phenomenon is analogous to the
resonance in dynamical systems 
whose parameters periodically vary with time -- parametric resonance. It was 
therefore called parametric resonance of neutrino oscillations \cite{ETC,Akh1}. 
In particular, in \cite{Akh1} the parametric resonance was considered for neutrino
oscillations in a matter with a periodic step function (``castle wall'') density 
profile, which allows an exact analytic solution. 

The parametric resonance can lead to large probabilities of neutrino flavor 
transition in matter even if the mixing angles both in vacuum and in matter 
are small. 
This happens because each half-wave oscillation of the transition 
probability is placed on the top of the previous one, {\em i.e.} 
the transition probability ``drifts up''. 
If mixing angle in matter is 
very small (matter density is far from the MSW resonance density), the parametric 
resonance enhancement of neutrino oscillations can manifest itself only if the 
neutrinos pass through a large number of periods of density modulation, {\it i.e.}  
travel a sufficiently long distance. However, if matter density is not very 
far from the MSW resonance one, an interesting interplay between the MSW effect 
and parametric effects can occur. In particular, a strong parametric enhancement 
of neutrino oscillations can take place even if the neutrinos pass only through 
1 - 2 periods of density modulation \cite{KS}.
 
Although the parametric resonance of neutrino oscillations is certainly an 
interesting physical phenomenon, it requires that very special conditions 
be satisfied. 
Unfortunately, these conditions cannot be created in the laboratory because 
this would require either too long a baseline or neutrino propagation 
in a matter of too high a density (see Sec. 5 below). 
Until recently it was also unclear whether a natural object exists where these
conditions can be satisfied for any known source of neutrinos. 
This situation has changed with a very important observation by Liu and Smirnov 
\cite{LS} (see also \cite{LMS}), who have shown that the parametric resonance 
conditions can be approximately satisfied for the oscillations of atmospheric 
$\nu_\mu$ into sterile neutrinos $\nu_s$ inside the earth. 

It is known that the earth consists of two main structures -- the mantle
and the core. Within the mantle and within the core the matter density changes 
rather slowly (the density variation scale is large compared to the typical
oscillation lengths of atmospheric and solar neutrinos), but at their border it 
jumps sharply by about a factor of two. Therefore to a good approximation one 
can consider the mantle and the core as structures of constant densities equal 
to the corresponding average densities (two-layer model). 
Neutrinos coming to the detector from the lower hemisphere at zenith
angles $\Theta$ in the range defined by
$\cos \Theta=(-1) \div (-0.837)$ traverse the earth's mantle, core and then
again mantle. Therefore such neutrinos experience a periodic ``castle wall'' 
potential, and their oscillations can be parametrically enhanced. Even though 
the neutrinos pass only through ``1.5 periods'' of density modulations (this 
would be exactly one period and a half if the distances neutrinos travel 
in the mantle and in the core were equal), the parametric effects on neutrino
oscillations in the earth can be quite strong. Subsequently, it has been pointed 
out by Petcov \cite{P1} that the parametric resonance conditions can also be 
satisfied (and to even a better accuracy) for the 
$\nu_2\leftrightarrow \nu_{e}$ oscillations in the earth in the case of the 
$\nu_e$ - $\nu_{\mu(\tau)}$ mixing 
\footnote{In \cite{P1} a different name for this phenomenon was suggested --
neutrino oscillation length resonance. We prefer to follow the original terminology 
of \cite{ETC,Akh1}.}. 
This, in particular, may have important implications 
for the solar neutrino problem. The parametric resonance in the oscillations 
of solar and atmospheric neutrinos in the earth was further explored in a number 
of papers \cite{Akh2,ADLS,CMP}. In the present paper we continue the study of this 
phenomenon. The following issues are addressed: the stability of the resonance with 
respect to the variation of the zenith angle of neutrino trajectory (Sec. 3), 
parametric resonances of higher order (Sec. 4), and the prospects of
experimental observation of the parametric resonance in neutrino oscillations 
(Sec. 5). In the last section we comment on a recent controversy regarding the 
physical nature of the resonance enhancement of the oscillations of the core 
crossing neutrinos in the earth. 

\section{Parametric resonance in neutrino oscillations in the earth}
In this section we shall summarize the main formulas describing neutrino
oscillations in the earth which will be relevant for the subsequent discussion. 
We will be essentially following the formalism and notation of ref. \cite{Akh2}. 

Consider the oscillations in a 2-flavor neutrino system in a matter with
periodic step function density profile. We will be assuming that one period of
density modulation consists of two parts of the lengths $R_m$ and $R_c$, 
with the corresponding effective matter densities $N_m$ and 
$N_c$ (``castle wall'' density profile). For the $\nu_e \leftrightarrow 
\nu_{\mu(\tau)}$ oscillations the effective matter density coincides with the
electron number density, whereas for the $\nu_{e,\mu,\tau} \leftrightarrow
\nu_{s}$ oscillations in an isotopically symmetric matter it is a factor of two 
smaller. The parametric resonance in such a system occurs when the oscillations
phases $2\phi_m$ and $2\phi_c$ acquired over the intervals $R_m$ and $R_c$ are 
odd integer multiples of $\pi$ \cite{Akh1,LS,LMS}. Let us denote 
\be
\delta=\frac{\Delta m ^2}{4E}\,,\quad\quad V_i=\frac{G_F}{\sqrt{2}}\,
N_i \,,\quad\quad 
\omega_i=\sqrt{(\cos 2\theta_{0}\,\delta-V_i)^2+
(\sin 2\theta_{0}\,\delta)^2}\,\quad\quad
(i=m,~c)\,.
\label{not}
\ee
Here $E$, $\Delta m^2$ and $\theta_0$ are the neutrino energy, mass squared
difference and vacuum mixing angle, respectively. The difference of the neutrino 
eigenenergies in a matter of density $N_i$ is $2\omega_i$, so that the oscillations 
phases acquired over the intervals $R_m$ and $R_c$ are 
\be
2\phi_m = 2\omega_m R_m\,,\quad\quad  2\phi_c = 2\omega_c R_c\,.  
\label{phases}
\ee
The parametric resonance conditions can be written as 
\cite{ETC,Akh1,KS,LS,LMS,P1,Akh2}
\footnote{In refs. \cite{ETC,Akh1,KS} these conditions were derived for the 
particular case $k=k'$, which, however, includes the most important principal 
resonance with $k=k'=0$.}
\be
\phi_1=\frac{\pi}{2}+k\pi\,, \quad\quad \phi_2=\frac{\pi}{2}+k'\pi\,,
\quad\quad k, k'=0,1,2,...
\label{rescond}
\ee
The evolution of a system of oscillating neutrinos is conveniently described 
by the evolution matrix $U$. In the case of the 2-flavor system $U$ is a 
$2\times 2$ unitary matrix which can be parametrized as 
\be
U
=Z-i\mbox{\boldmath $\sigma$}{\bf W}\,, \quad\quad Z^2+{\bf W}^2=1\,.
\label{U3}
\ee
Here $\mbox{\boldmath $\sigma$}$ are the Pauli matrices in the flavor space.   
The evolution matrix $U$ describes the evolution of an arbitrary initial state 
and therefore contains all the information relevant for neutrino oscillations.  
In particular, the probabilities of the neutrino flavor oscillations $P$ and of 
$\nu_2 \leftrightarrow \nu_{e}$ oscillations $P_{2e}$ are 
given by \cite{Akh2} 
\be
P=W_1^2+W_2^2\,,\quad\quad 
P_{2e}=\sin^2 \theta_0+W_1(W_1 \cos 2\theta_0+W_3 \sin 2\theta_0)\,.
\label{prob}
\ee

We shall consider neutrino oscillations in the earth in the two-layer 
approximation. 
Neutrinos coming to the detector from the lower hemisphere of the earth at 
zenith angles $\Theta$ in the range $\cos \Theta=(-1) \div (-0.837)$ 
(nadir angle $\Theta_n \equiv 180^\circ - \Theta \le 33.17^\circ$) traverse 
the earth's mantle, core and then again mantle, {\it i.e.} three layers of 
constant density with the third layer being identical to the first one. In 
this case the parameters $Z$, ${\bf W}$ take a very simple form \cite{Akh2}: 
\be
Z=2 \cos\phi_m \,Y-\cos\phi_c \,,
\label{Z}
\ee
\be
{\bf W}=\left( 2 \sin\phi_m \sin 2\theta_m \,Y+\sin\phi_c \sin 2\theta_c \,, ~0\,,
~-\left(2 \sin\phi_m \cos 2\theta_m \,Y + \sin\phi_c \cos 2\theta_c\right)
\right) \,.
\label{W}
\ee
Here $\theta_m$ and $\theta_c$ are the neutrino mixing angles in the matter of 
the density $N_m$ and $N_c$ ({\it i.e.} in the mantle and in the core of the 
earth), 
\be
Y=\cos\phi_m \cos\phi_c-\sin\phi_m \sin\phi_c \cos(2\theta_m - 2\theta_c) \,,  
\label{Y}
\ee
and the vector ${\bf W}$ was written in components. At the parametric resonance, 
{\it i.e.} when the conditions (\ref{rescond}) are satisfied, the neutrino flavor
transition probability takes the value \cite{LS,LMS},  
\be
P=\sin^2 (2\theta_c-4\theta_m) \,,
\label{prob1}
\ee
whereas the probability of the $\nu_2 \leftrightarrow \nu_{e}$ transitions is 
\cite{P1} 
\be
P_{2e} = \sin^2 (2\theta_c-4\theta_m+\theta_0) \,.
\label{prob2}
\ee
These probabilities can be close to unity (the arguments of the sines 
close to $\pi/2$) even if the amplitudes of neutrino oscillations in the mantle, 
$\sin^2 2\theta_m$, and in the core, $\sin^2 2\theta_c$, are rather small. 
This can happen if the neutrino energy lies in the range $E_c < E < E_m$, 
where $E_m$ and $E_c$ are the values of the energy that correspond to the 
MSW resonance in the mantle and in the core of the earth.  
The probability $P_{2e}$ is relevant for the description of the oscillations 
of solar neutrinos in the earth \cite{MS2,BW1}. In the case of small mixing
angle MSW solution of the solar neutrino problem, $\sin^2 2\theta_0 < 10^{-2}$ 
\cite{BKS}, 
and $P_{2e}$ practically coincides with $P$ unless both probabilities are very
small.  

The trajectories of neutrinos traversing the earth are determined by their 
nadir angle $\Theta_n$.  
The distances $R_m$ and $R_c$ that neutrinos travel in the mantle (each layer) and
in the core are given by 
\be
R_m=R\left(\cos\Theta_n-\sqrt{r^2/R^2-\sin^2\Theta_n}\,\right)\,,
\quad\quad R_c=2R \,\sqrt{r^2/R^2-\sin^2\Theta_n}\,.
\label{RmRc}
\ee
Here $R=6371$ km is the earth's radius and $r=3486$ km is the radius of the core. 
The matter density in the mantle of the earth ranges from 2.7 $g/cm^3$ at the
surface to 5.5 $g/cm^3$ at the bottom, and that in the core ranges from 
9.9 to 12.5 $g/cm^3$ (see, {\it e.g.}, \cite{Stacey}). The electron number 
fraction $Y_e$ is close to 1/2 both in the mantle and in the core.
Taking the average matter densities in the mantle and core to be 4.5 and 11.5
$g/cm^2$ respectively, one finds for the $\nu_e\leftrightarrow \nu_{\mu,\tau}$
oscillations involving only active neutrinos the following values of $V_m$
and $V_c$: $V_m=8.58\times 10^{-14}$ eV, $V_c=2.19\times 10^{-13}$ eV.
For transitions involving sterile neutrinos $\nu_e\leftrightarrow
\nu_{s}$ and $\nu_{\mu,\tau}\leftrightarrow \nu_s$, these parameters
are a factor of two smaller.

\section{Stability of the parametric resonance conditions}
If the parametric resonance conditions (\ref{rescond}) are satisfied, strong 
parametric enhancement of the oscillations of core crossing neutrinos in the earth
can occur \cite{LS,LMS,P1,Akh2,ADLS,CMP} 
\footnote{For the parametric resonance to be a maximum of the transition
probability, an additional condition has to be satisfied 
(two conditions in the case of $\nu_2 \leftrightarrow \nu_{e}$ oscillations
\cite{P1}). For the oscillations between neutrinos of different flavor, this 
condition can be written as  
$\cos(2\theta_c-4\theta_m)<0$. 
For small vacuum mixing angles $\theta_0$ it essentially reduces to 
$\delta <V_c$, with a small region around $\delta=V_m$ excluded.}. 
We shall now discuss these conditions. The phases $\phi_m$ and $\phi_c$ 
depend on the neutrino parameters $\Delta m^2$, $\theta_0$ and $E$ and also on 
the distances $R_m$ and $R_c$ that the neutrinos travel in the mantle and in the
core. The path lengths $R_m$ and $R_c$ vary with the nadir angle; however, as 
can be seen from (\ref{RmRc}), their changes are correlated and they cannot take 
arbitrary values. Therefore if for some values of the neutrino parameters 
a value of the nadir angle $\Theta_n$ exists for which, for example, the first
condition in eq. (\ref{rescond}) is satisfied, it is not obvious if at the same 
value of $\Theta_n$ the second condition will be satisfied as well. In other 
words, it is not clear if the parametric resonance conditions can be fulfilled 
for neutrino oscillations in the earth for at least one set of the neutrino 
parameters $\Delta m^2$, $\theta_0$ and $E$. However, as was shown in 
\cite{P1,Akh2}, not only the parametric resonance conditions are satisfied (or 
approximately satisfied) for a rather wide range of the nadir angles covering 
the earth's core, they are fulfilled for the ranges of neutrino parameters 
which are of interest for the neutrino oscillations solutions of the solar and
atmospheric neutrino problems. In particular, the conditions for the principal 
resonance ($k=k'=0$) are satisfied to a good accuracy for $\sin^2 2\theta_0 
\aprle 0.1$, $\delta \simeq (1.1 \div 1.9)\times 10^{-13}$ eV$^2$, which includes  
the ranges relevant for the small mixing angle MSW solution of the solar neutrino
problem and for the subdominant $\nu_\mu \leftrightarrow \nu_{e}$ 
and $\nu_e \leftrightarrow \nu_{\tau}$ oscillations of atmospheric neutrinos. 

The fact that the parametric resonance conditions can be satisfied so well for 
neutrino oscillations in the earth is rather surprising. It is a consequence 
of a number of remarkable numerical coincidences. It has been known for some 
time \cite{GKR,LS,LiLu} that the potentials $V_m$ and $V_c$   
corresponding to the matter densities in the mantle and core, the inverse
radius of the earth $R^{-1}$, and typical values of $\delta\equiv
\Delta m^2/4E$ of interest for solar and atmospheric neutrinos, are all of
the same order of magnitude -- ($3\times 10^{-14}$ -- $3\times 10^{-13}$) eV.
It is this surprising coincidence that makes appreciable earth effects on the 
oscillations of solar and atmospheric neutrinos possible.
However, for the parametric resonance to take place, a coincidence 
by an order of magnitude is not sufficient: the conditions 
(\ref{rescond}) have to be satisfied at least within a 50\% accuracy \cite{Akh2}. 
This is exactly what takes place. In addition, in a wide range of the nadir 
angles $\Theta_n$, with changing $\Theta_n$ the value of the parameter $\delta$ 
at which the resonance conditions (\ref{rescond}) are satisfied slightly 
changes, but the fulfillment of these conditions is not destroyed. We shall show 
now that in this row of mysterious coincidences, at least the last one -- 
the stability of the parametric resonance conditions with respect to variations 
of the nadir angle -- has a simple explanation. 

We begin with simple qualitative arguments. Consider the principal parametric 
resonance for which $k=k'=0$. In figs. 1 and 2 the flavor transition 
probability $P$ and the parameters $\phi_m-\pi/2$ and $\phi_c-\pi/2$ are plotted as 
functions of $\delta$ for $\sin^2 2\theta_0=0.01$ and $\Theta_n=27.3^\circ$. The
highest peak of $P$ corresponds to $\delta \simeq 1.55 \times 10^{-13}$ eV at which, 
as can be seen in fig. 2, both $\phi_m-\pi/2$ and $\phi_c-\pi/2$ nearly cross the 
abscissa, {\it i.e.} the conditions (\ref{rescond}) are approximately satisfied. 
This is the 
parametric resonance peak; the smaller peaks on its left and right are due to the
MSW resonances in the mantle and in the core, respectively. The minima of the curves 
$\phi_m-\pi/2$ and $\phi_c-\pi/2$ correspond to the minima of $\omega_m$ and 
$\omega_c$.  
The values of $R_m$ and $R_c$ determine the steepness of the curves. With 
increasing nadir angle, $R_m$ increases and $R_c$ decreases, which means that 
the branches of the $\phi_m-\pi/2$ curve become steeper and those of the 
$\phi_c-\pi/2$ curve become less steep. Increasing the steepness of 
$\phi_m-\pi/2$ would push the intersection point of the $\phi_m-\pi/2$ and 
$\phi_c-\pi/2$ curves upwards; however, decreasing the steepness of $\phi_c-\pi/2$  
would push this intersection point in the opposite direction. As a result, 
in a wide range of the nadir angles $\Theta_n$, with changing $\Theta_n$ the 
intersection point shifts along the abscissa but remains close to it, {\it i.e.}  
the parametric resonance conditions are not destroyed. It is easy to see that 
this only happens when the intersection takes place between the minima of 
$\phi_m-\pi/2$ and $\phi_c-\pi/2$, {\it i.e.} for $\delta$ in the range 
\be
V_m/\cos 2\theta_0 < \delta <V_m/\cos 2\theta_0 \,. 
\label{interval}
\ee
If the intersection takes place outside this interval, the changes of the steepness 
of the $\phi_m-\pi/2$ and $\phi_c-\pi/2$ curves due to changing $\Theta_n$ will push 
their intersection point in the same direction, and this will quickly destroy the 
parametric resonance conditions. Therefore outside the range (\ref{interval}) the 
parametric resonance can only occur in rather narrow intervals of $\Theta_n$. 
Interestingly, the interval (\ref{interval}) is not only favored due to the wide 
range of allowed values of $\Theta_n$, but also corresponds to the region where the 
strongest parametric enhancement of neutrino oscillations takes place \cite{P1,Akh2}. 

We shall now quantify the above arguments. Each of the two parametric resonance
conditions in (\ref{rescond}) can be solved with respect to the value 
$\delta_0$ of the parameter $\delta$ 
at the resonance: 
\begin{eqnarray}
\delta_0 &=& \cos 2\theta_{0}\,V_m \pm \sqrt{[(2k+1)\pi]^2/4 R_m^2-\sin^2
2\theta_{0}\,V_m^2}
\label {delta1} \\
\delta_0 &=& \cos 2\theta_{0}\,V_c \pm \sqrt{[(2k'+1)\pi]^2/4 R_c^2-\sin^2   
2\theta_{0}\, V_c^2}
\label{delta2}
\end{eqnarray}
The parametric resonance can occur if the values of $\delta_0$ given by 
these two expressions coincide or nearly coincide. Notice that 
the existence of the radicals in (\ref{delta1}) and (\ref{delta2}) 
in general puts upper limits on the vacuum mixing angle $\theta_0$. In particular, 
for the principal resonance, 
assuming that the resonance condition can 
be satisfied for all neutrino trajectories that cross the core (including the
vertical ones), from (\ref{delta2}) one finds \cite{Akh2} 
\be
\sin^2 2\theta_{0}\le \frac{\pi^2}{4(R_c)_{max}^2\,V_c^2}\simeq 0.04\,.
\label{limit}
\ee
If one excludes small nadir angles, the constraint becomes less stringent. For
example, for $\sin^2 \Theta_n \ge 0.12$ one obtains $\sin^2 2\theta_{0}\le 0.07$. 
The analogous upper bound following from (\ref{delta1}) are less restrictive except 
for the trajectories close to the edge of the core of the earth. 

The ($\pm$) signs in eqs. (\ref{delta1}) and (\ref{delta2}) correspond to different 
regimes of the parametric resonance. We shall now choose the plus sign in 
(\ref{delta1}) and the minus sign in (\ref{delta2}), which for small $\theta_0$ 
corresponds to the interval (\ref{interval}). The other cases can be considered 
similarly 
\footnote{In the cases that correspond to the two plus signs, the parametric 
resonance does not lead to a maximum of the transition probability; for two 
minus signs there exist parametric maxima, but they are typically smaller than 
in the region of $\delta$ that we are considering.}. 
  
The consistency of eqs. (\ref{delta1}) and (\ref{delta2}) requires 
\be
\cos 2\theta_{0}\,(V_c-V_m) =\sqrt{\frac{[(2k+1)\pi]^2}{4 R_m^2}-\sin^2
2\theta_{0}\,V_m^2}\;+\;\sqrt{\frac{[(2k'+1)\pi]^2}{4 R_c^2}-\sin^2 2\theta_{0}\,
V_c^2}\,. 
\label{consist1}
\ee
For small vacuum mixing angles $\theta_0$ this simplifies to 
\footnote{For the particular case $k=k'=0$ this condition was obtained in 
\cite{P1}.}
\be
(V_c-V_m) =\frac{\pi}{2}\left(\frac{2k+1}{R_m}+\frac{2k'+1}{R_c}\right)\,.
\label{consist2}
\ee
Eq. (\ref{consist1}) was obtained from the requirement that both conditions in 
(\ref{rescond}) are satisfied at the same value of $\delta$. If it is fulfilled,   
with changing $\Theta_n$ the resonance value of $\delta$ changes, but the parametric 
resonance conditions are still satisfied. Thus, it is the condition (\ref{consist1})
that ensures the stability of the parametric resonance conditions with respect to 
the variations of the nadir angle. It remains now to check if or when this condition 
is indeed satisfied. We shall do that in the limit of the small $\theta_0$ when 
the simplified form (\ref{consist2}) applies. Similar analysis can be performed 
with the use of eq. (\ref{consist1}). 

The l.h.s. of eq. (\ref{consist2}) is $\Theta_n$ independent, so must be the 
r.h.s. As we have pointed out above, with increasing $\Theta_n$ the path length
$R_m$ increases whereas $R_c$ decreases, which is just a consequence of the 
spherical geometry of the earth. One can therefore expect at least a partial 
cancellation of the changes of the $(2k+1)/R_m$ and $(2k'+1)/R_c$ terms in their 
sum on the r.h.s. of eq. (\ref{consist2}). We shall show now that for the principal 
parametric resonance with $k=k'=0$ this cancellation is indeed quite good in a
rather wide range of $\Theta_n$. In fig. 3 the nadir angle dependence of $R_m^{-1}$, 
$R_c^{-1}$ and $(R_m^{-1}+R_c^{-1})$ is plotted. One can see that while $R_m^{-1}$ 
and $R_c^{-1}$ change noticeably with $\Theta_n$, their sum remains almost constant 
in the interval $\Theta_n=0 \div 0.4$. For larger values of $\Theta_n$ this sum 
increases, but the increase does not exceed 25\% for $\Theta_n \aprle 0.54$. 
Thus, if the parametric resonance conditions are satisfied at some value of
$\Theta_n$ in the interval $0\le \Theta_n \aprle 0.54$, they will be approximately 
satisfied in the whole interval. This explains the stability of the parametric
resonance conditions. The fact that they {\em are} satisfied to a good accuracy 
for at least one value in the above-mentioned interval is a consequence of the
numerical coincidences that we discussed earlier. 

\section{Higher-order parametric resonances}
Eq. (\ref{consist1}) can be used for a systematic search for the parametric
resonances of neutrino oscillations in the earth. Once a solution is found 
({\it i.e.} the values of $R_m$ and $R_c$ are determined), one can find the 
corresponding value of $\delta_0$ from (\ref{delta1}) or (\ref{delta2}). 
The properties of the principal resonance have been discussed in some detail 
in \cite{LS,LMS,P1,Akh2}; here we shall concentrate on the higher-order 
resonances.

The resonances of interest are located in or close to the region (\ref{interval}). 
The resonances that are far from that region either do not correspond to maxima or
lead to small values of the transition probability. In addition, they can exist 
only in rather narrow intervals of the nadir angle. We have been able to identify 
two higher-order resonances of the $\nu_e \leftrightarrow \nu_{\mu(\tau)}$ 
oscillations. Let us discuss their properties and possible implications.

(i) The resonance with $k=0$, $k'=1$. The resonance enhancement of the
transition probability occurs in the interval of the nadir angles $\Theta_n 
\simeq 20^\circ \div 25^\circ$ and for relatively large vacuum mixing angles, 
$\sin^2 2\theta_0 \simeq 0.68 \div 0.86$. It corresponds to $\delta_0\simeq 
(7 \div 9)\times 10^{-14}$ eV, depending on the values of $\Theta_n$ and 
$\sin^2 2\theta_0$. For typical values of these parameters the resonance is 
illustrated by figs. 4 and 5. One can see that there is practically no detuning, 
{\it i.e.} the conditions (\ref{rescond}) are satisfied very well. The parametric
peak at $\delta_0 \simeq 9\times 10^{-14}$ eV is high and very wide, the energy 
width at half height being $\Delta E/E_0 \sim 1$ (the energies $E_1$ and $E_2$ at 
the low-energy and high-energy borders of the half-height region satisfy $E_2/E_1 
\aprge 3$). The parametric resonance conditions can also be fulfilled for 
$\sin^2 2\theta_0 < 0.68$, but in this region the resonance corresponds 
to a low-lying saddle point rather than to a maximum of the transition probability. 
The existence of the resonance with $k=0$, $k'=1$ was pointed out in \cite{CMP},
but its properties were not discussed there. 

(ii) The resonance with $k=0$, $k'=2$. 
This resonance occurs for small values of the nadir angle, 
$\Theta_n \aprle 12^\circ$, and large (close to maximal) vacuum mixing, 
$\sin^2 2\theta_0 \aprge 0.9$. The resonance values of $\delta$ lie 
in the range $\delta_0\simeq (0.65 \div 1.1)\times 10^{-13}$ eV. With 
increasing $\sin^2 2\theta_0$ the resonance value of $\delta$ decreases.  
This resonance is shown for $\Theta_n=0$ and $\sin^2 2\theta_0=0.95$ in 
figs. 6 and 7. The parametric peak at $\delta_0\simeq 1 \times 10^{-13}$ eV is 
slightly less wide than in the case (i): $E_2/E_1 \simeq 2.1$. One can 
see from fig. 6 that the curves  $\phi_m-\pi/2$  and $\phi_c-5\pi/2$ cross the  
abscissa at different but close points which are located symmetrically around 
$\delta=1\times 10^{-13}$ eV. The detuning at $\delta=1\times 10^{-13}$ eV 
does not exceed $\pm 0.15$, {\it i.e.} is rather small. 

What are possible implications of these higher-order resonances? Unlike the 
principal resonance which can occur for both small and large values of
$\theta_0$, the resonances (i) and (ii) are only possible if the vacuum mixing 
angle is rather large, $\sin^2 2\theta_0\aprge 0.7$. For atmospheric
neutrinos, for values of $\Delta m^2$ of practical interest, such large mixing 
in the $\nu_e-\nu_\mu$ system is excluded by the Super-Kamiokande 
\cite{SK1} and CHOOZ \cite{CHOOZ} data. Large mixing angle is in principle 
of interest for the vacuum oscillations and large mixing angle MSW  solutions of 
the solar neutrino problem. However, the former requires very small values of 
$\delta$, orders of magnitude below those at which the parametric resonance 
of the neutrino oscillations in the earth can occur. The large mixing angle MSW
solution, on the contrary, requires too large values of $\delta$. For $\Delta m^2 
\simeq 7\times 10^{-6}$ eV$^2$ which is the smallest value allowed by the total 
event rates of the four solar neutrino experiments \cite{BKS}, and $E=10$ MeV 
which is a typical boron neutrino energy, one gets $\delta\simeq 1.75 \times
10^{-13}$ eV. 
This is about a factor of 1.6 - 2.5 larger than the values of $\delta_0$ 
that are necessary for the higher-order parametric resonances discussed above. 
These resonances could, however, occur for $hep$ neutrinos
which constitute the highest-energy component of the solar neutrino flux (endpoint 
energy 18.8 MeV). Smaller values of $\Delta m^2$ (and, consequently, of $\delta$) are 
allowed if the boron neutrino flux is considered as a free parameter \cite{BKS,Barb}.  
Relatively small values of $\Delta m^2$ are also allowed in 3-flavor analyses
of the solar neutrino data, even when the constraints on the mixing matrix 
element $U_{e3}$ following from the CHOOZ experiments are taken into account
\cite{FLM1,Barb}. Notice, however, that although the values $\Delta m^2 \simeq 
7\times 10^{-6}$ eV$^2$ or even less are allowed at large vacuum mixing angles by
the total event rates, they are disfavored by the recent Super-Kamiokande data on 
the zenith angle distribution of the solar neutrino events.

We conclude that the higher-order parametric resonances are less interesting than 
the principal one. The domain of the neutrino parameters in which they can occur 
for the oscillations of solar neutrino in the earth is currently disfavored by 
the solar neutrino data, though not completely ruled out. These resonances are 
of no interest for the atmospheric neutrino oscillations. It would be interesting 
to look for other possible neutrino sources for which these resonances could play 
a role. 

\section{Can the parametric resonance of neutrino oscillations be observed?}
Besides being an interesting physical phenomenon, the parametric resonance of 
neutrino oscillations can provide us with an important additional information  
about neutrino properties. 
Therefore experimental observation of this effect would be of considerable 
interest. We shall now discuss the prospects for experimental observation 
of the parametric resonance of neutrino oscillations in the earth, having 
in mind mainly the principal resonance. There are two main sources of neutrinos 
for which the parametric resonance can be important -- atmospheric neutrinos 
and solar neutrinos. Both sources have their advantages and disadvantages 
from the point of view of the possibility of observation of the parametric 
resonance. We shall now briefly discuss them. 

We start with atmospheric neutrinos. The parametric resonance can occur in the 
$\nu_\mu \leftrightarrow \nu_s$ \cite{LS,LMS} and also in the subdominant 
$\nu_e \leftrightarrow \nu_{\mu(\tau)}$ channels of oscillations \cite{ADLS}.  
It can affect the distributions of $\mu$-like events and also (in the case of 
the $\nu_e \leftrightarrow \nu_{\mu(\tau)}$ oscillations) lead to interesting 
peculiarities in the zenith angle distributions of the multi-GeV e-like events. 

The observation of the parametric effects is hampered 
by the loose correlation between the directions of the momenta of atmospheric
neutrinos and of the charged leptons which they produce and which are actually 
detected. Because of this the trajectories of neutrinos coming to the detector 
are not known very precisely. In addition, the data are presented for certain 
samples of events (sub-GeV, multi-GeV, upward through-going, upward stopping) 
which includes collecting data over rather wide energy intervals. The 
contributions of the parametric peaks may therefore be integrated over together 
with other possible enhancement peaks -- due to the MSW resonances in the mantle 
and in the core, making the distinction between these effects difficult. Also, 
strong resonance enhancement effects (both parametric and MSW) can only occur 
either for neutrinos or for antineutrinos, depending on the sign of $\Delta m^2$ 
\footnote{In general, the parametric enhancement can take place even for ``wrong 
sign'' $\Delta m^2$, but in the case of neutrino oscillations in the earth 
these effects are small.}. 
The present atmospheric neutrino experiments do not distinguish between 
neutrinos and antineutrinos, therefore possible matter effects are ``diluted'' 
in the sum of the $\nu$- and $\bar{\nu}$-induced events. At certain values of 
the ratio of the muon and electron neutrino fluxes $r(E,\Theta_n)$ depending on 
the value of mixing angle $\theta_{23}$ the parametric effects on e-like events 
are suppressed \cite{ADLS}. 

Atmospheric neutrinos have some advantages for observation of the parametric 
resonance of neutrino oscillations in the earth. Neutrinos come to the detectors 
from all directions, which means that practically the whole solid angle covering 
the earth's core will contribute to the effect. There are no additional suppression 
factors due to a specific composition of the incoming  neutrino flux which may
quench the earth's matter effect on the oscillations of solar neutrinos (see below). 
Parametric effects may provide a unique probe of the neutrino mixing angle
$\theta_{13}$ with sensitivity going far beyond that of the long-baseline 
accelerator and reactor experiments \cite{ADLS,ALS}. 
Possible ways of improving the prospects of the experimental observation of the 
parametric effects in the atmospheric neutrino oscillations include using
various energy cuts, finer zenith angle binning and detectors capable of detecting 
the recoil nucleon, which would enable one to reconstruct the direction of an
incoming neutrino \cite{ADLS,ALS}. 

Solar neutrinos can experience a strong parametric enhancement of their oscillations 
in the earth if the small mixing angle MSW effect is the correct explanation 
of the the solar neutrino deficit \cite{P1,Akh2}. The parametric enhancement can 
occur in a wide range of values of $\sin^2 2\theta_0$ and for the nadir angles 
$\Theta_n$ almost completely covering the core of the earth. The trajectory of 
each detected neutrino is exactly known. For boron neutrinos the resonance occurs 
at the values of $\Delta m^2$ which correspond to the central part of the allowed
interval for the small mixing angle MSW solution of the solar neutrino problem. 

However, there are some disadvantages, too. 
Unfortunately, due to their geographical  location, the existing solar
neutrino detectors have a relatively low time during which
solar neutrinos pass through the core of the earth to reach the detector
every calendar year. The Super-Kamiokande detector has a largest fractional
core coverage time equal to $7\%$. In \cite{GKR} it was suggested to build
a new detector close to the equator in order to increase the sensitivity   
to the earth regeneration effect; this would also maximize the parametric
resonance effects in oscillations of solar neutrinos in the earth.
In the case of the MSW solutions of the solar neutrino problem the probability 
$P_{SE}$ of finding a solar $\nu_e$ after it traverses the earth depends 
sensitively on the average $\nu_e$ survival probability in the sun $\bar{P}_S$ 
\cite{MS2,BW1}:
\be
P_{SE}=\bar{P}_S + \frac{1-2 \bar{P}_S}{\cos 2\theta_0}\,(P_{2e}-\sin^2 \theta_0)\,.
\label{PSE}
\ee 
The probability $P_{2e}$ can experience a strong parametric enhancement, but 
in the case of small mixing angle MSW solution of the solar neutrino problem  
the probability $\bar{P}_S$ for the Super-Kamiokande and SNO experiments 
turns out to be rather close to 1/2. This means that the effects of passage 
through the earth on solar neutrinos should be strongly suppressed. The 
current best fit of the solar neutrino data is not far from the line  
in the parameter space where $\bar{P}_S$ is exactly equal to 1/2 and 
$P_{SE}=\bar{P}_S$ ({\it i.e.} the earth matter effects are absent). Whether or 
not it will be possible to observe the parametric resonance of the oscillations of 
solar neutrinos in the earth depends on how close to this line the true values of 
$\sin^2 2\theta_0$ and $\Delta m^2$ are. By now the Super-Kamiokande experiment  
has not observed, within its experimental accuracy, any enhancement of neutrino 
signal for earth core crossing neutrinos \cite{SK2}. This can be because 
the parametric enhancement of the neutrino oscillations in the 
earth does not occur ({\it e.g.} if the true solution of the solar neutrino 
problem is vacuum oscillations or large mixing angle MSW effect), or because 
the values of $\Delta m^2$ and $\sin^2 2\theta_0$ are too close to those at which 
$\bar{P}_S=1/2$. Hopefully, with accumulated statistics of the Super-Kamiokande and 
forthcoming data from the SNO experiment the situation will soon be clarified. 

It is interesting to note that the Super-Kamiokande data on the zenith angle 
dependence of the solar neutrino events seems to indicate some deficiency 
of the events due to the core-crossing neutrinos rather than an excess  
\cite{SK2}, although it is not statistically significant. Should this 
deficiency be confirmed by future data with better statistics, it could have 
a natural explanation in terms of the parametric resonance of neutrino 
oscillations. As follows from (\ref{PSE}), the parametric enhancement of 
$P_{2e}$ for core crossing neutrinos can lead to a deficiency of the events 
if the neutrino parameters are in the small-$\sin^2 2\theta_0$ part of the 
allowed region which corresponds to $P_S>1/2$ (see, {\it e.g.}, fig. 10 in 
ref. \cite{BK}). In this case one should also have an ``opposite sign'' overall 
day-night effect (fewer events during the night than during the day). 
In any case, given the current experimental constraints 
on the neutrino parameters, if the small mixing angle MSW effect is the true 
solution of the solar neutrino problem, the only hope to observe earth matter 
(day-night) effects on solar neutrinos seems to be through the parametric 
resonance of oscillations of core crossing neutrinos.

As we have seen, observing the parametric resonance in oscillations of solar 
and atmospheric neutrinos in the earth is not an easy task. Can one create 
the necessary matter density profile and observe the parametric resonance in 
neutrino oscillations in the laboratory ({\it i.e.} short-baseline) experiments?  
Unfortunately, the answer to this question seems to be negative. The parametric 
resonance can occur when the mean oscillation length in matter approximately 
coincides with the matter density modulation length \cite{ETC,Akh1,KS}: 
$l_m\simeq L$. In a matter of density $N_i$ the oscillation length is given by
$l_m=\pi/\omega_i$ where $\omega_i$ was defined in (\ref{not}). Let us require 
$l_m \aprle 1$ km. Assume first that $V_i\aprge\delta$, {\it i.e.} $\omega_i$ are 
dominated by matter density terms. Than for $l_m\aprle 1$ km one would need 
a matter of mass density $\rho_i\ge 3.3\times 10^4$ g/cm$^3$, clearly not a 
feasible value. Conversely, for $\rho_i\le 10$ g/cm$^3$, one finds $l_m\aprge 
3300$ km, a distance comparable with the earth's radius. Consider now the opposite 
case, $\delta \gg V_i$. Then the oscillation length in matter essentially coincides 
with the vacuum oscillations length which in principle can be rather short provided 
that the vacuum mixing angle $\theta_0$ is small (otherwise this would contradict 
reactor and accelerator data). However, in this case there is another problem. 
Requiring $l_m\aprle 1$ km one finds $\delta \aprge 2.5 \times 10^{-10}$ eV. For 
$\rho_i\aprle 10$ g/cm$^3$ one therefore has $V_i/\delta \aprle 10^{-3}$. This means 
that the mixing angles in matter are very close to the vacuum one, $\theta_i\simeq 
\theta_0 (1+V_i/\delta)$, and so their difference is very small: $\Delta \theta =
\theta_2-\theta_1 \simeq (\Delta V/\delta)\,\theta_0\aprle 10^{-3}\theta_0$.  
When the difference of mixing angles in matter is small, the 
parametric effects can manifest themselves only if neutrinos travel over a large 
number of periods, $n\simeq \pi/4 \Delta\theta$ \cite{ETC,Akh1,KS,Akh2}. Therefore 
in this case the necessary baseline is $\sim \pi^2/(4\theta_0 \Delta V) \aprge 
3\times 10^3$ km, again too large. One can conclude that the sole presently known 
object where the parametric resonance of neutrino oscillations can take place 
is the earth, as was first pointed out in \cite{LS,LMS}.

\section{Parametric resonance or oscillation length resonance?}
Recently, there have been a considerable controversy regarding the physical 
nature of the resonance enhancement of the oscillations of the core crossing
neutrinos in the earth. In refs. \cite{LS,LMS,Akh2,Sm1,Sm2,ADLS} this enhancement  
was interpreted as a parametric resonance effect. However, in refs. 
\cite{P1,P2,P3,CMP} it was interpreted as a new effect, called neutrino 
oscillation length resonance.  
It was argued that this effects has nothing to do with the parametric resonance of 
neutrino oscillations. This controversy has led to a considerable confusion and 
we therefore believe that it is worthwhile to discuss the issue in some detail.  
We shall argue here that the resonance enhancement of 
oscillations of core crossing neutrinos is a typical parametric resonance effect.  

The general possibility of the parametric resonance in neutrino oscillations 
was suggested independently in \cite{ETC} and \cite{Akh1} and further studied 
in \cite{KS}. In these papers it was shown that the parametric
resonance is a specific resonance enhancement of neutrino transition 
probability which occurs due to a certain relationship between the 
density modulation scale and neutrino oscillation length, does not require 
large amplitude of neutrino oscillations, and has a property to accumulate 
when the matter density profile repeats itself, {\it i.e.} is periodic. 
The resonance enhancement of the oscillations of core crossing neutrinos 
in the earth fully matches this description. 
The conditions (11) in \cite{P1} (reproduced also in \cite{P2,P3,CMP,P4}), 
which are claimed to be the conditions for the oscillation length resonance, 
coincide with the parametric resonance conditions given in 
\cite{ETC,Akh1,KS,LS,LMS} (see eqs. (10) and (17) in \cite{Akh1}, eqs. (2) and (4) 
in \cite{KS}), and in particular eq. (7) in \cite{LS} and eq. (12) in \cite{LMS} 
where the parametric resonance conditions were formulated specifically for the case 
of the oscillations of core crossing neutrinos in the earth).
Moreover, the probability of the neutrino flavor transition 
at the oscillation length resonance given in \cite{CMP} coincides with 
the oscillation probability at the parametric resonance given in 
\cite{LS,LMS} (see eq. (\ref{prob1}) above). 

We shall now comment on the arguments presented in refs. \cite{P1,P2,P3,P4}  
(and given below in italics) in support of the point of view that the resonance
enhancement of the oscillations of core crossing neutrinos in the earth is very 
different from the parametric resonance effects in the case of the ``castle wall'' 
matter density profile studied in \cite{Akh1}.

\noindent
$\bullet$ {\it In \cite{Akh1} the results were presented for the specific 
case of small densities ($V_{m,c}\ll \delta$) which does not correspond to the  
situation with oscillations of neutrinos in the earth. }

The case $V_{m,c}\ll \delta$ was discussed in ref. \cite{Akh1} just in order 
to stress that the proposed effect -- parametric resonance in neutrino 
oscillations -- is very different from the MSW effect. The parametric resonance 
can take place for any relation between the matter-induced potentials and 
$\delta$ (see, {\it e.g.}, \cite{KS}). 

\noindent
$\bullet$ {\it In \cite{Akh1} the case of the density profile 
with equal spatial dimensions of the layers of constant density ($R_m=R_c$) 
was studied, whereas in the earth $R_m\ne R_c$. }

The nature of the effect does not depend on whether $R_m$ and $R_c$ are equal 
to each other or not. In fact, there is a value of the nadir angle ($\Theta_n=
27.3^\circ$) which determines the neutrino trajectory through the earth
along which $R_m=R_c$. Fig. 1 was drawn for this particular value of the nadir
angle; it shows a rather typical pattern of the resonance enhancement in the 
oscillations of core crossing neutrinos (compare with analogous figures in 
refs. \cite{P1,Akh2,ADLS,CMP}
\footnote{Similar figures can also be found in some earlier papers 
\cite{numer} the authors of which have found the parametric resonance of 
neutrino oscillations in the earth in numerical calculations but have not 
recognized its parametric nature.}
). 

\noindent
$\bullet$ {\em The density profile along the trajectory of core crossing
neutrinos is not periodic even in the two-layer approximation; neutrino path 
length falls short even of 1.5 periods. }

We shall first comment on the second part of this argument. In general, it is 
incorrect: neutrino path length is less than 1.5 periods of density modulation 
for $\cos\Theta_n >0.89$ but exceeds this value for  
$0.837<\cos\Theta_n <0.89$, which constitutes about 31\% of the solid angle 
covering the earth's core. This, however, is unimportant. The term ``1.5 periods'' 
was used in refs. \cite{Akh2,ADLS} just as a shorthand for describing the 
situation when neutrinos cross three layers of constant density with the third 
layer being identical to the first one. 

In fig. 8 we show the coordinate dependence of the probability of neutrino 
oscillations in the earth for $\Theta_n=11.5^\circ$  
(solid curve), and for comparison also show the probability in the hypothetical 
case when neutrinos propagate through the full two periods of density modulation 
(dashed curve). Obviously, in both cases the physical nature of the enhancement 
of the transition probability is the same: the probability 
drifts up and each new half-wave piece of the oscillating curve is placed on 
the top of the previous one. This pattern is typical of the parametric resonance 
of neutrino oscillations \cite{ETC,Akh1,Akh2}. 

Now about the periodicity of the matter density profile of the earth in the 
two-layer approximation. In the strict mathematical sense, a function $f(x)$ 
can be called periodic only if it is defined in the infinite interval 
$x\in (-\infty,\infty)$. It was shown, however, in \cite{ETC,Akh1} that the 
complete flavor conversion is possible even if neutrinos travel over a finite 
distance in a matter of periodic density. In \cite{KS} it was demonstrated 
that this distance can be quite short (1-2-3 periods) if the parametric 
resonance energy is not very far from the MSW resonance energy. This is exactly 
what happens in the case of the oscillations of the core crossing neutrinos 
in the earth. The term ``periodic'' was used in refs. \cite{ETC,Akh1,KS,Akh2} 
in this limited sense -- as a finite piece of the periodic density profile.
Since the path length of core crossing neutrinos in the earth exceeds one  
period of density modulation $R_m+R_c$, the density profile of the earth in 
the two-layer approximation satisfies this definition.

Actually, the only relevant question here is if it is really important that 
the third layer traversed by the neutrinos going through the earth is 
identical to the first one. This brings us to the next objection. 

\noindent
$\bullet$ {\em The specific enhancement of the neutrino transition
probability under discussion can take place even if neutrinos 
cross a number of layers of constant but different densities. }

In general, this is correct. Assume, for example, that the neutrinos cross 
three layers of densities $N_1<N_2<N_3$, and that we are free to choose 
the spatial extensions of these layers, so that the condition similar to 
(\ref{rescond}) can be satisfied. Then, in principle, one can have a strong 
enhancement of the transition probability. However, the resonance conditions 
(\ref{rescond}) are just the extremality conditions; for the resonance to be
a maximum of the transition probability, additional conditions have to be 
satisfied. In the case of neutrino flavor oscillations in a step-function 
density profile with $N_3=N_1$, there is only one such condition; in the 
case $N_1<N_2<N_3$ there are three additional conditions. Thus, in this 
latter case there are altogether six conditions (three conditions on oscillation 
phases and three additional conditions), whereas in the case $N_3=N_1$ 
there are only three conditions. 

As was discussed above, the possibility of the resonance enhancement of 
the oscillations of the core crossing neutrinos in the earth is a result of 
a number of remarkable numerical coincidences. It is because of these 
coincidences that all three conditions for the parametric resonance of neutrino 
oscillations can be satisfied. It is much less probable that {\em six} 
different conditions could be satisfied as a result of some coincidences 
of parameters. Fortunately, in the case of the oscillations of the core 
crossing neutrinos in the earth, the third layer that the neutrinos cross 
{\em is} identical to the first one, {\it i.e.} the matter density profile 
{\em is} periodic and so there is a chance that the resonance enhancement of 
neutrino oscillations under the discussion will be observed.

This work was supported by Funda\c{c}\~ao para a Ci\^encia e a Tecnologia 
through the grant PRAXIS XXI/BCC/16414/98 and also in part by the TMR network 
grant ERBFMRX-CT960090 of the European Union.

\newpage
\centerline{\large Figure captions}

\vglue 0.4cm
\noindent
Fig. 1. 
Transition probability $P$ for $\nu_e\leftrightarrow \nu_{\mu,\tau}$ 
oscillations in the earth for core-crossing
neutrinos vs $\delta$. $\sin^2 2\theta_{0}=0.01$, $\Theta_n=27.3^\circ$.

\noindent
Fig. 2. 
Phases $\phi_m-\pi/2$ (dashed curve) and $\phi_c-\pi/2$ (solid curve) vs
$\delta$ for $\nu_e\leftrightarrow \nu_{\mu,\tau}$ oscillations in the earth. 
$\sin^2 2\theta_{0}=0.01$, $\Theta_n=27.3^\circ$. 

\noindent
Fig. 3. 
Parameters $1/R_m$ (long-dashed curve), $1/R_c$ (short-dashed curve) and 
$1/R_m+1/R_c$ (solid curve) as functions of the nadir angle $\Theta_n$.
All distances are in the units of the earth's radius $R$; $\Theta_n$ is
in radians.

\noindent
Fig. 4. 
Same as fig. 1 but for $\sin^2 2\theta_{0}=0.77$, 
$\Theta_n=23.3^\circ$.  

\noindent
Fig. 5. 
Phases $\phi_m-\pi/2$ (solid curve) and $\phi_c-3\pi/2$ (dashed curve) vs
$\delta$ for $\nu_e\leftrightarrow \nu_{\mu,\tau}$ oscillations in the earth. 
$\sin^2 2\theta_{0}=0.77$, $\Theta_n=23.3^\circ$.  

\noindent
Fig. 6. 
Same as fig. 1 but for $\sin^2 2\theta_{0}=0.95$, 
$\Theta_n=0$.

\noindent
Fig. 7. 
Phases $\phi_m-\pi/2$ (solid curve) and $\phi_c-5\pi/2$ (dashed curve) 
vs $\delta$ for $\nu_e\leftrightarrow \nu_{\mu,\tau}$ 
oscillations in the earth. $\sin^2 2\theta_{0}=0.95$, $\Theta_n=0$.

\noindent
Fig. 8. 
Solid curve: transition probability $P$ for 
$\nu_e\leftrightarrow \nu_{\mu,\tau}$ oscillations 
in the earth as a function of the distance $t$ (measured in units of
the earth's radius $R$) along the neutrino trajectory. $\delta=1.8\times 
10^{-13}$ eV, $\sin^2 2\theta_{0}=0.01$, $\Theta_n=11.5^\circ$. Dashed 
curve: the same for a hypothetical case of neutrino propagation over 
full two periods of density modulation ($t_{max}=2(R_m+R_c$)).

\newpage
\begin{figure}[H]
\vglue -3cm 
\mbox{\epsfig{figure=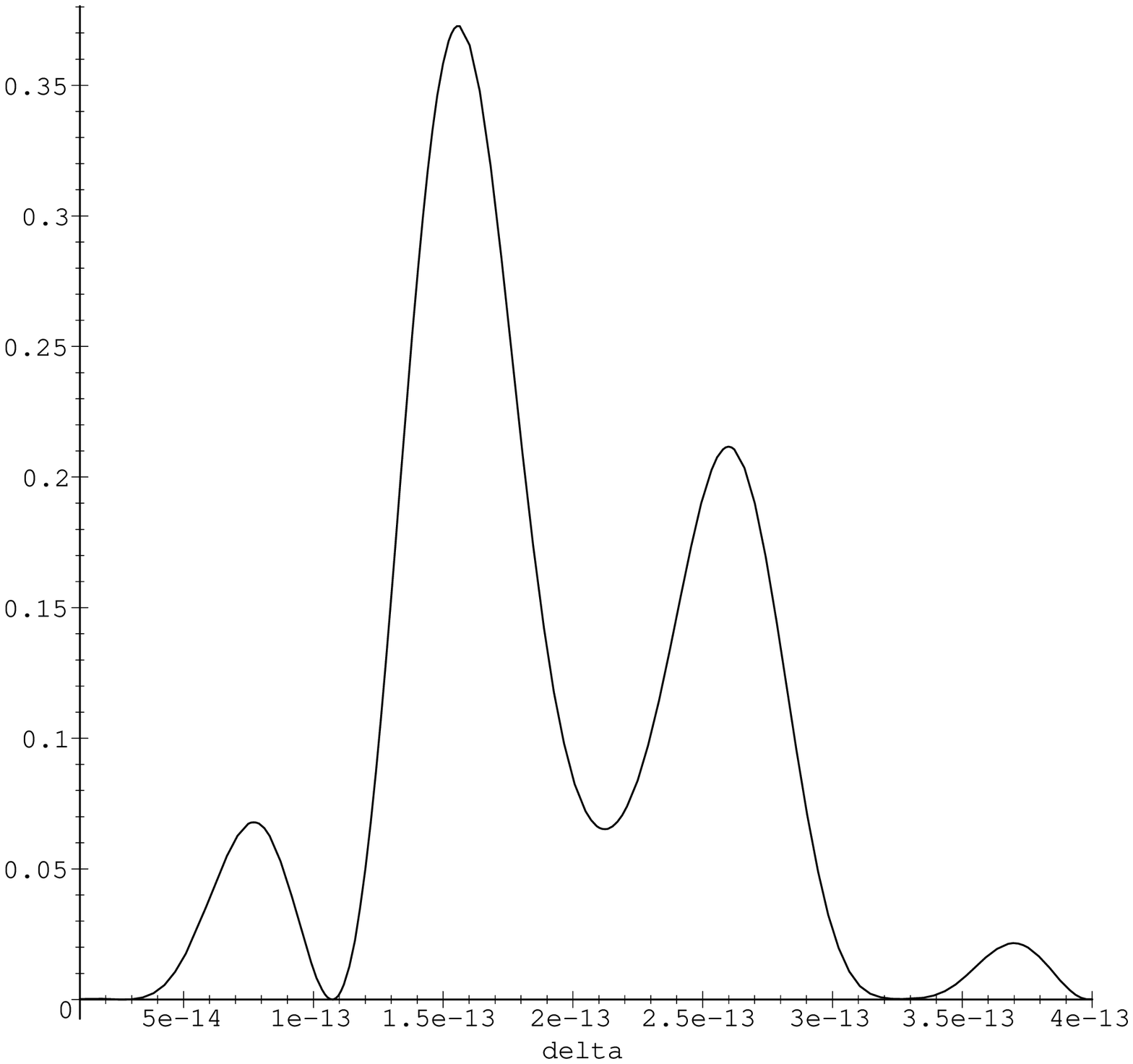,width=15.5cm, height=12.5cm}}
\vglue -1.0cm
\centerline{\mbox{Fig. 1.}}
\mbox{\epsfig{figure=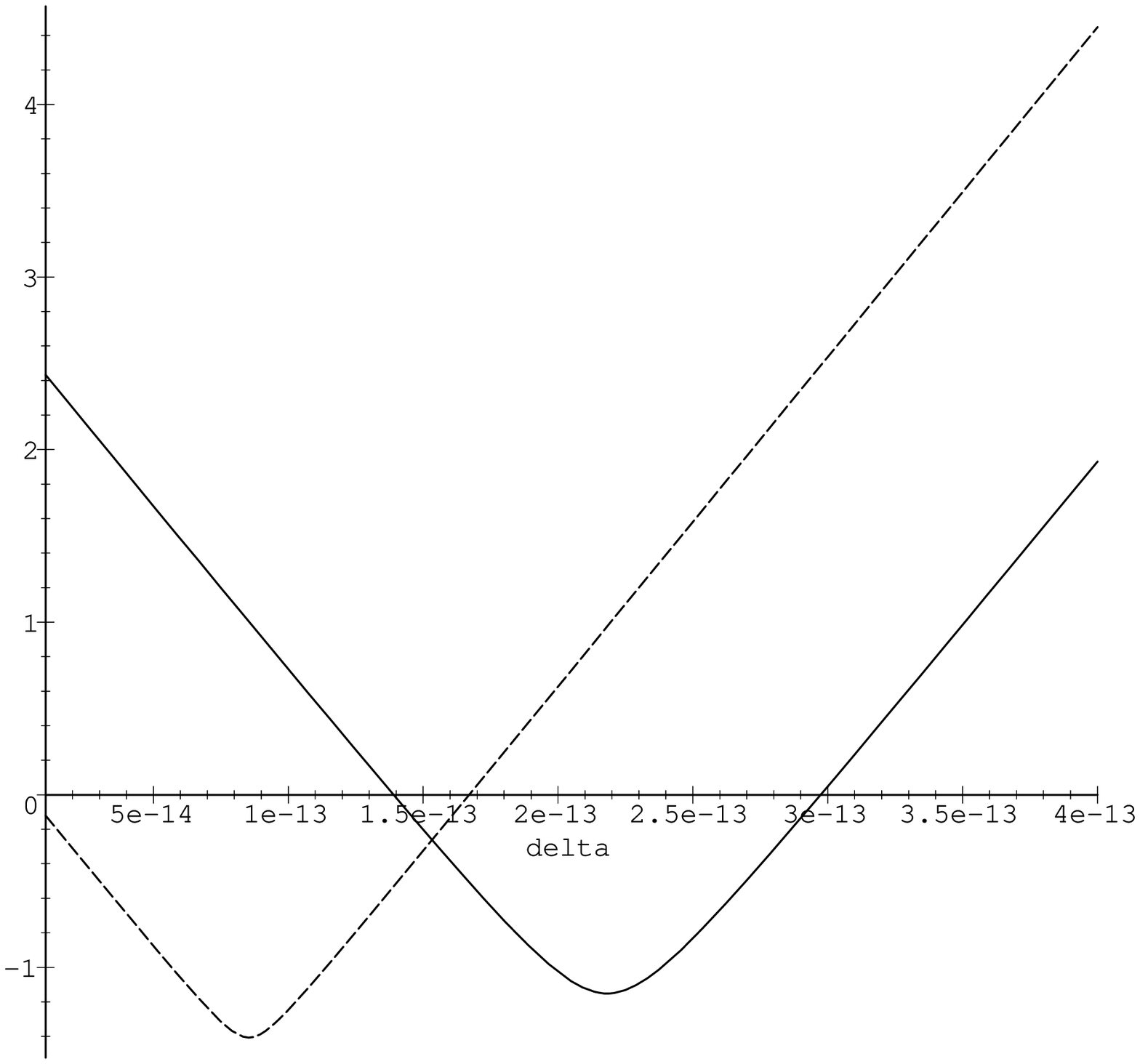,width=15.5cm, height=12.5cm}}
\vglue -1.5cm
\centerline{\mbox{Fig. 2.}}
\end{figure}

\begin{figure}[H]
\hglue -2.0cm
\mbox{\epsfig{figure=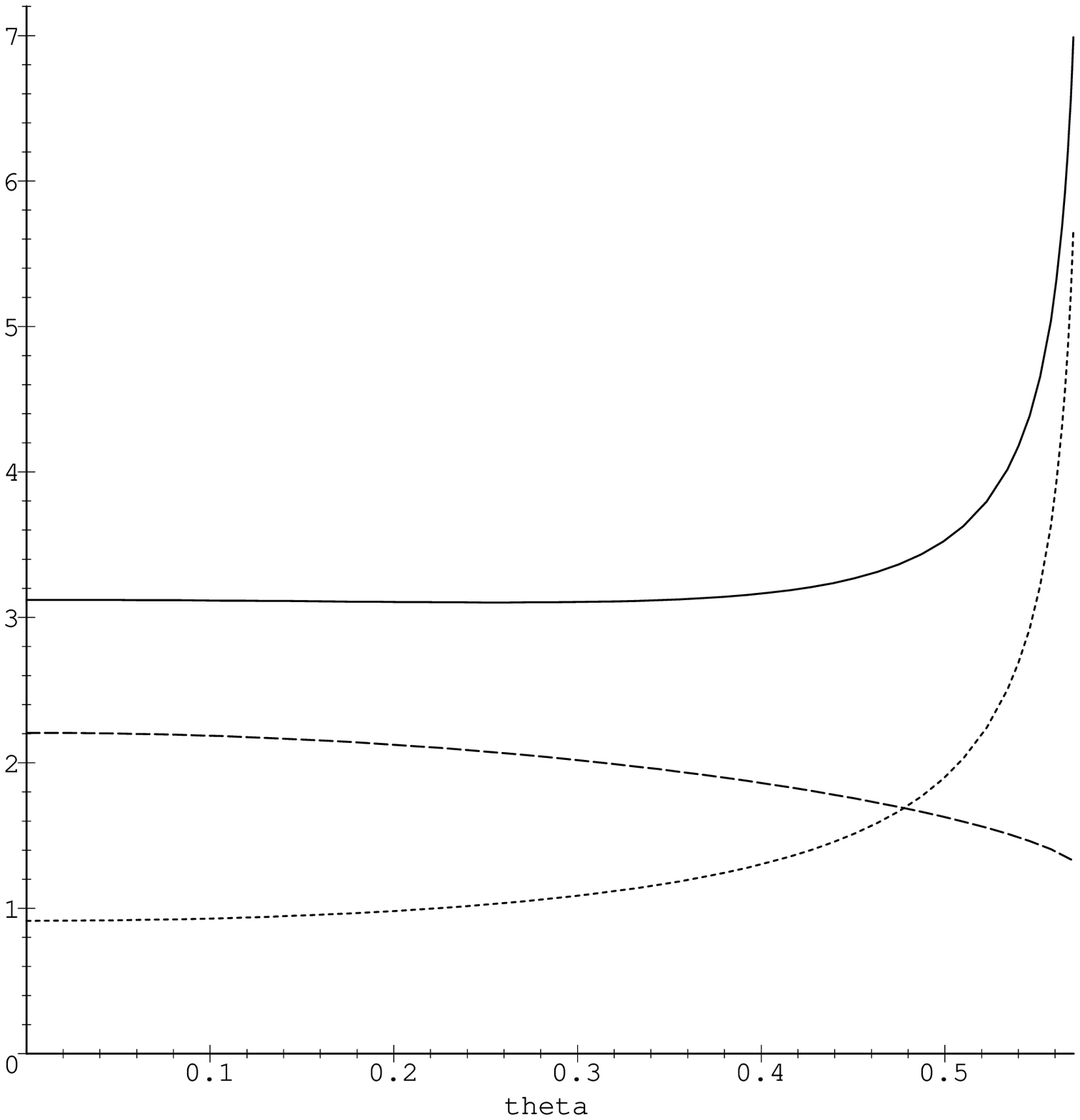,width=19.0cm, height=19.0cm}}
\vglue -1.5cm
\centerline{\mbox{Fig. 3.}}
\end{figure}

\newpage
\begin{figure}[H]
\vglue -3cm 
\mbox{\epsfig{figure=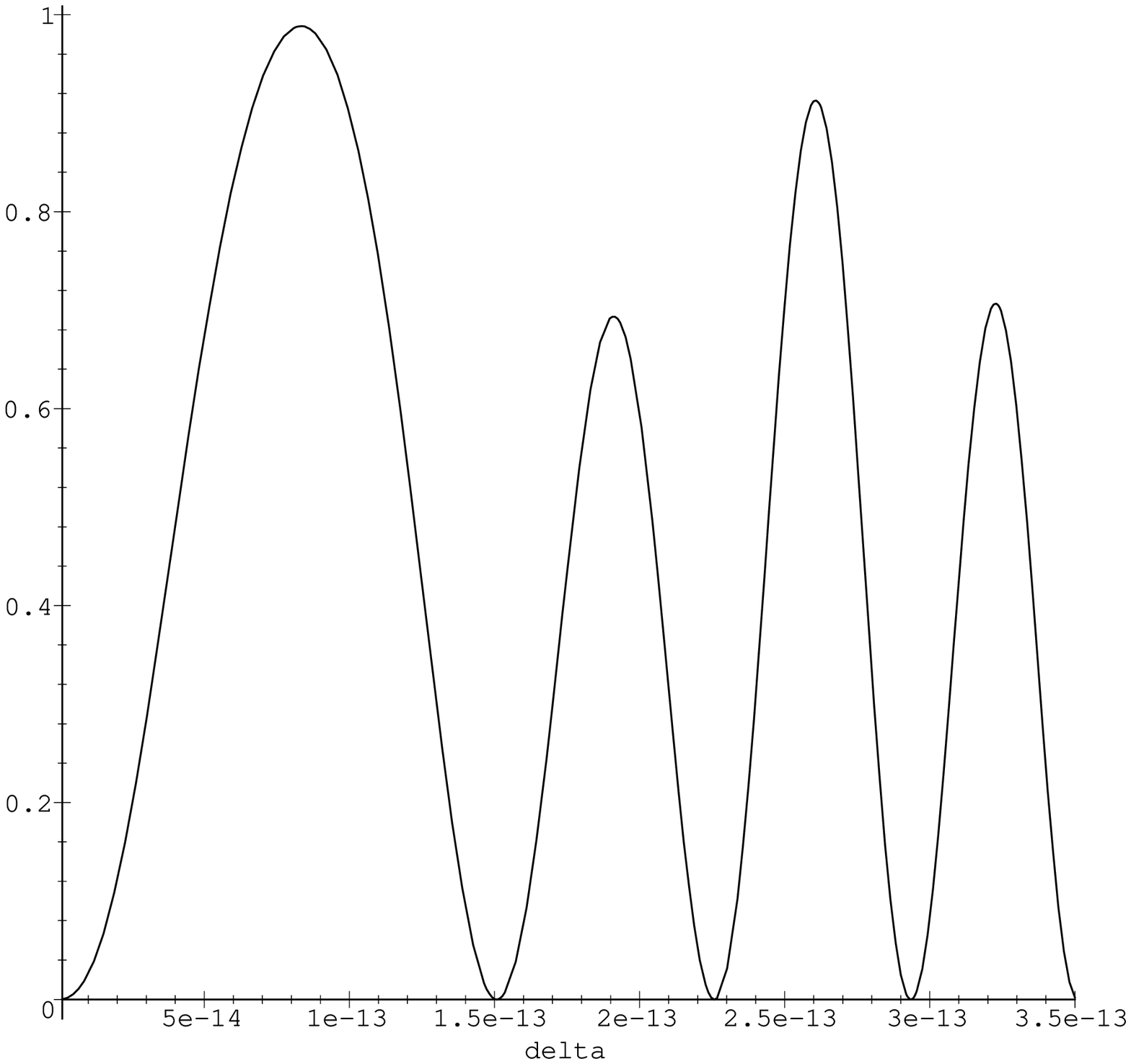,width=15.5cm, height=12.5cm}}
\vglue -1.0cm
\centerline{\mbox{Fig. 4.}}
\vglue -1cm
\mbox{\epsfig{figure=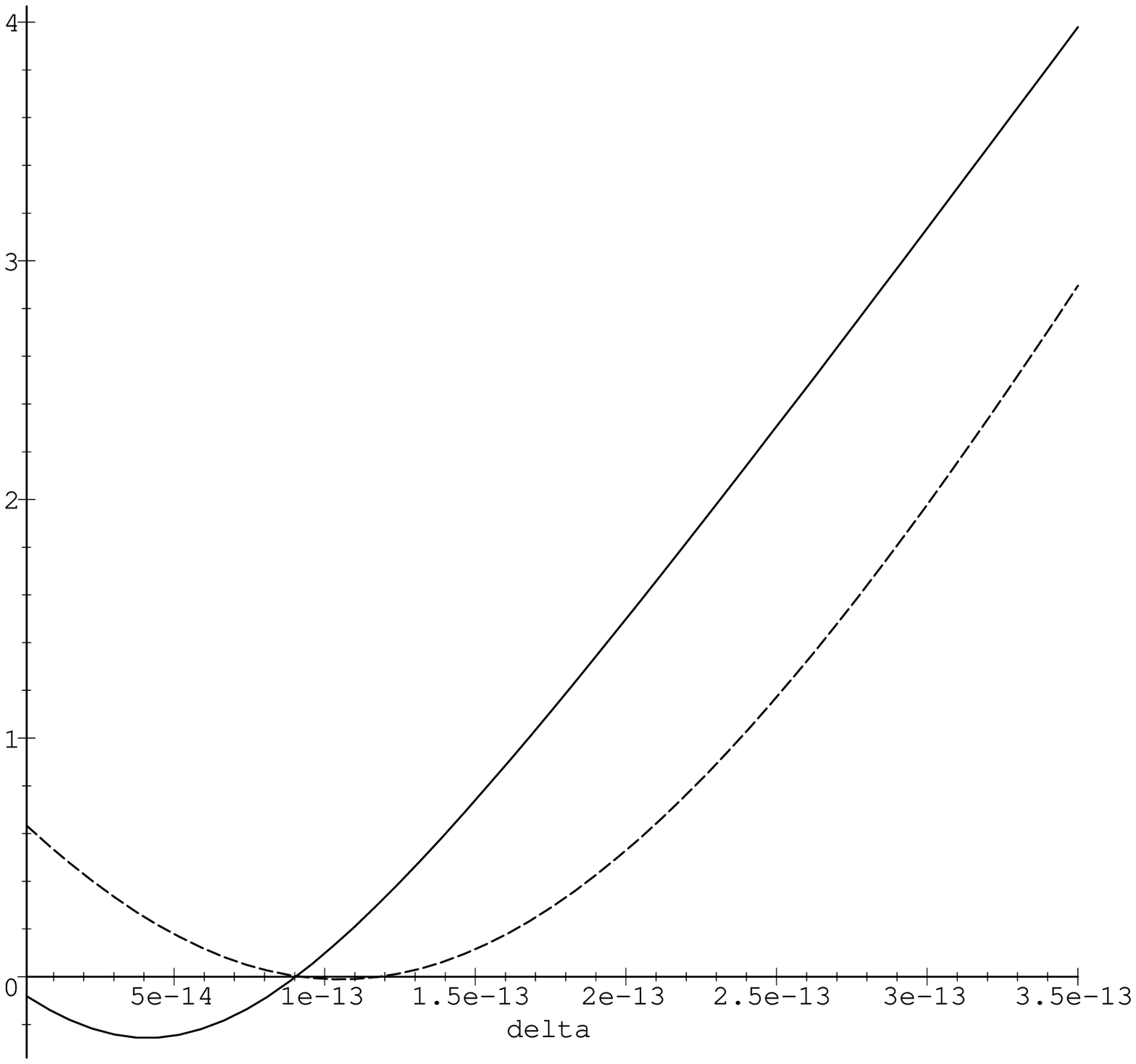,width=15.5cm, height=12.5cm}}
\vglue -1.5cm
\centerline{\mbox{Fig. 5.}}
\end{figure}

\newpage
\begin{figure}[H]
\vglue -3cm 
\mbox{\epsfig{figure=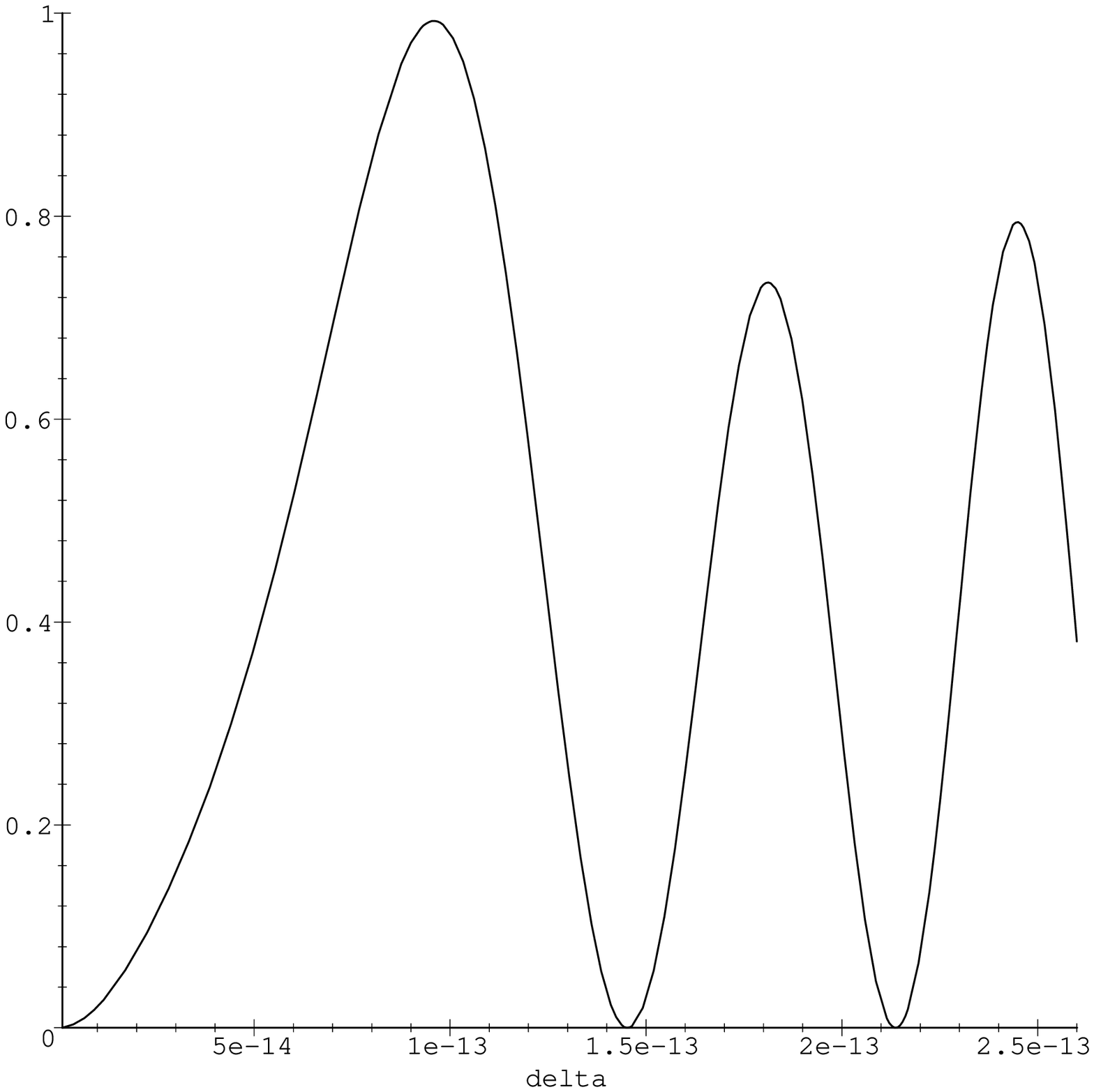,width=15.5cm, height=12.5cm}}
\vglue -1.0cm
\centerline{\mbox{Fig. 6.}}
\mbox{\epsfig{figure=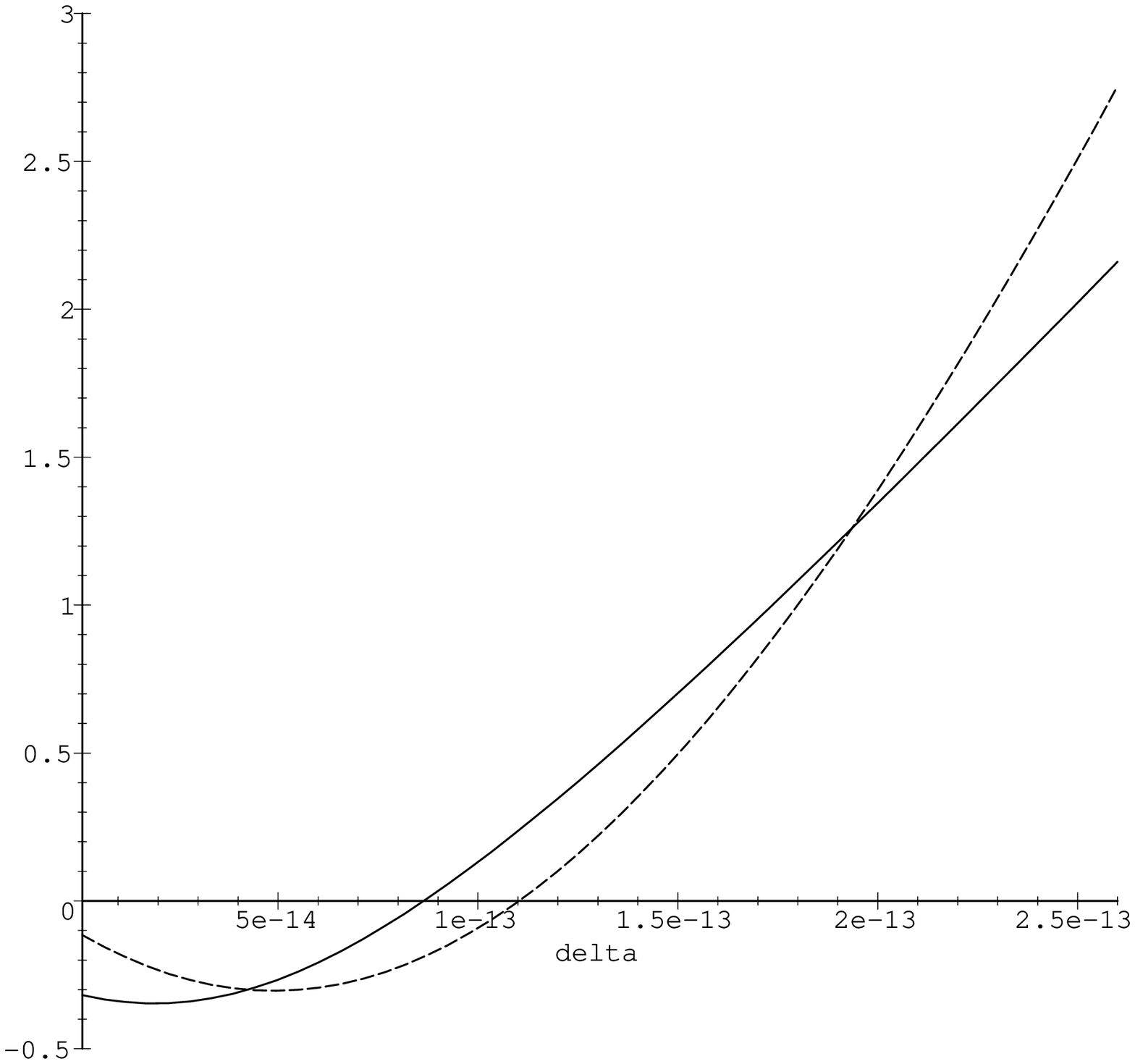,width=15.5cm, height=12.5cm}}
\vglue -1.5cm
\centerline{\mbox{Fig. 7.}}
\end{figure}


\begin{figure}[H]
\hglue -2.0cm
\mbox{\epsfig{figure=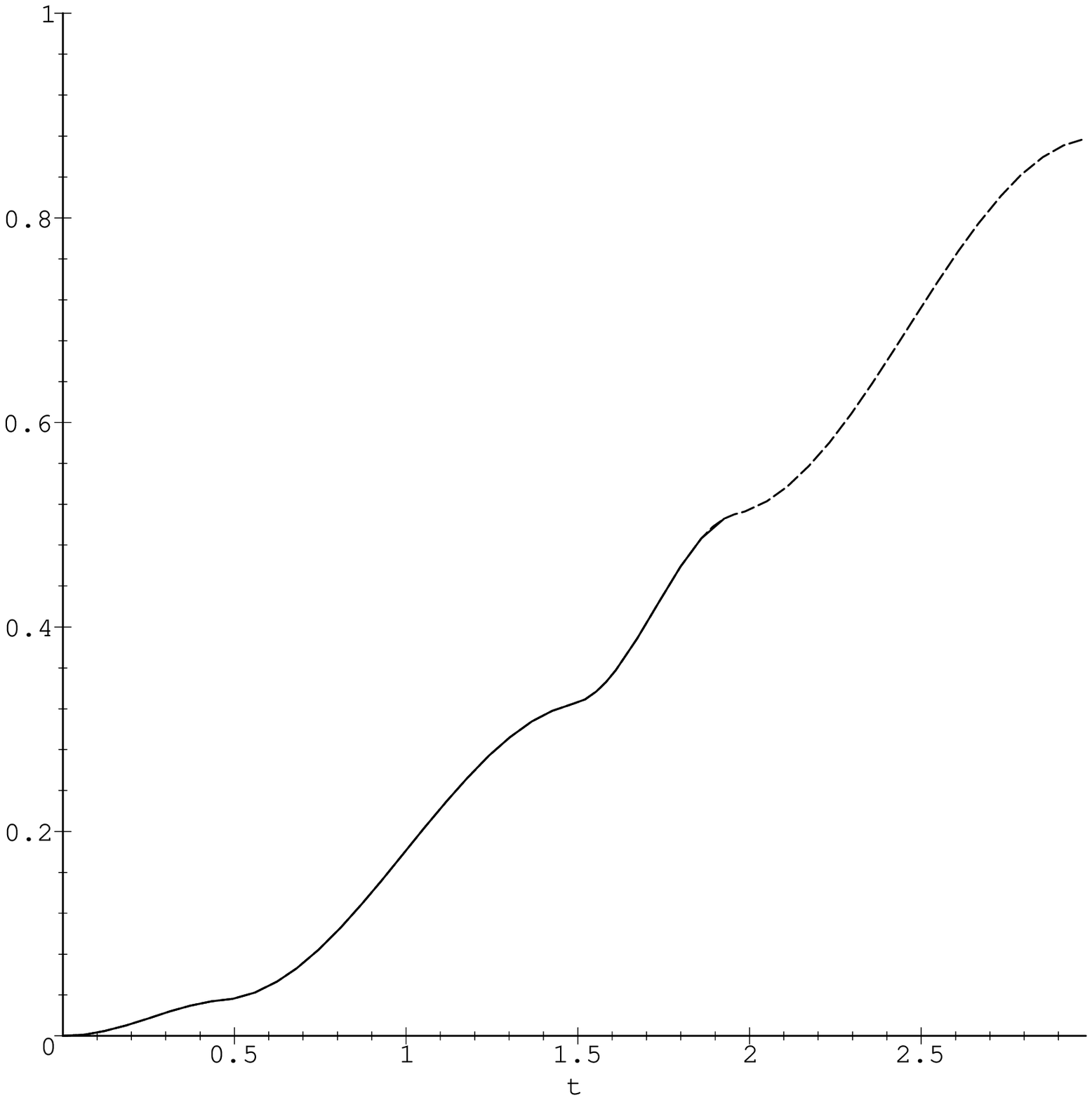,width=19.0cm, height=19.0cm}}
\vglue -1.5cm
\centerline{\mbox{Fig. 8.}}
\end{figure}


\begin{thebibliography}{99}
\bibitem{MS} S. P. Mikheyev and A. Yu. Smirnov, Sov. J. Nucl. Phys. 42 
(1985) 913.
\bibitem{W} L. Wolfenstein, Phys. Rev. D 17 (1978) 2369. 
\bibitem{ETC} V. K. Ermilova, V. A. Tsarev and V. A. Chechin, Kr. Soob,
Fiz. [Short Notices of the Lebedev Institute]  5 (1986) 26. 
\bibitem{Akh1} E. Kh. Akhmedov, preprint IAE-4470/1, 1987; Yad. Fiz. 
47 (1988) 475 [Sov. J. Nucl. Phys. 47 (1988) 301]. 
\bibitem{KS} P. I. Krastev and A. Yu. Smirnov, Phys. Lett. B 226 (1989) 341.  
\bibitem{LS} Q. Y. Liu and A. Yu. Smirnov, Nucl. Phys. B524 (1998) 505. 
\bibitem{LMS} Q. Y. Liu, S. P. Mikheyev and A. Yu. Smirnov, Phys. Lett. B440 
(1998) 319. 
\bibitem{P1} S. T. Petcov, Phys. Lett. B434 (1998) 321 (hep-ph/9805262). 
\bibitem{Akh2} E. Kh. Akhmedov, Nucl. Phys. B538 (1999) 25 (hep-ph/9805272). 
\bibitem{ADLS} E. Kh. Akhmedov, A. Dighe, P. Lipari and A. Yu. Smirnov,
hep-ph/9808270 (Nucl. Phys. B, in press).  
\bibitem{CMP} M. Chizhov, M. Maris and S. T. Petcov, hep-ph/9810501.
\bibitem{MS2} S. P. Mikheyev and A. Yu. Smirnov, in {\it '87 New and 
Exotic Phenomena}, proceedings of the 7th Moriond Workshop, edited by O.
Fackler and J. Tr\^an Thanh V\^an (Editions Fronti\`eres, Gif-sur-Yvette,
1987), p. 405. 
\bibitem{BW1} A. J. Baltz and J. Weneser, Phys. Rev. D 35 (1987) 528. 
\bibitem{BKS} J. N. Bahcall, P. I. Krastev and A. Yu. Smirnov, 
Phys.Rev. D58 (1998) 096016.  
\bibitem{Stacey} F. D. Stacey, {\it Physics of the Earth}, John Wiley
and Sons, New York, 1969. 
\bibitem{GKR} J. M. Gelb, W.-K. Kwong, and S. P. Rosen, Phys. Rev. Lett. 
78 (1997) 2296. 
\bibitem{LiLu} P. Lipari and M. Lusignoli, Phys. Rev. D58 (1998) 073005.
\bibitem{SK1} Super-Kamiokande Collaboration, Y. Fukuda {\em et al.},
Phys. Rev. Lett. 81 (1998) 1562; 
K. Scholberg, Talk given at the VIII International Workshop  {\em "Neutrino
Telescopes"}, Venice, Italy, February 23-26, 1999. 
\bibitem{CHOOZ} CHOOZ collaboration, M. Apollonio {\it et al.},
Phys. Lett. B420 (1998) 397. 
\bibitem{Barb} R. Barbieri {\em et al.}, hep-ph/9807235. 
\bibitem{FLM1}  G. L. Fogli, E. Lisi, A. Marrone and G. Scioscia, 
Phys. Rev. D59 (1999) 033001.  
\bibitem{ALS} E. Kh. Akhmedov, P. Lipari, A. Yu. Smirnov, in progress. 
\bibitem{SK2} Super-Kamiokande Collaboration, Y. Fukuda {\em et al.}, 
Phys. Rev. Lett. 82 (1999) 1810; K.~Inoue, Talk given at the VIII International 
Workshop  {\em "Neutrino Telescopes"}, Venice, Italy, February 23-26, 1999. 
\bibitem{BK} J. N. Bahcall and P. I. Krastev, Phys. Rev. C 56 
(1997) 2839.  
\bibitem{Sm1} A. Yu. Smirnov, Talk given at the 18th International Conference 
on Neutrino Physics and Astrophysics {\em ``Neutrino 98''}, Takayama, Japan, 
June 4-9, 1998,  hep-ph/9809481. 
\bibitem{Sm2} A. Yu. Smirnov, Talk given at the Symposium {\em ``New Era in
Neutrino Physics''}, Tokyo, Japan, June 11-12 1998, hep-ph/9811296. 
\bibitem{P2} S. T. Petcov, Talk given at the 18th International Conference on 
Neutrino Physics and Astrophysics {\em ``Neutrino 98''}, Takayama, Japan, 
June 4-9, 1998, hep-ph/9809587.  
\bibitem{P3} S. T. Petcov,  Talk given at the Symposium {\em ``New Era in
Neutrino Physics''}, Tokyo, Japan, June 11-12 1998, hep-ph/9811205. 
\bibitem{P4} S. T. Petcov, Talk given at the 17th International Workshop on Weak
Interactions and Neutrinos {\em (WIN99)}, Cape Town, South Africa,
January 24-30, 1999. 

\bibitem{numer} 
S. P. Mikheyev and A. Yu. Smirnov, in {\it '86 Massive Neutrinos in
Astrophysics and in Particle Physics}, 
proceedings of the 6th Moriond Workshop, edited by O. Fackler and
J. Tr\^an Thanh V\^an (Editions Fronti\`eres, Gif-sur-Yvette, 1986),
p. 355; E. D. Carlson, Phys. Rev. D 34 (1986) 1454; A. J. Baltz and J. Weneser, 
ref. \cite{BW1}; M. Cribier, W. Hampel, J. Rich, and D. Vignaud, 
Phys. Lett. B 182 (1986) 89; A. Dar, A. Mann, Y. Melina, and D. Zajfman, 
Phys. Rev. D 35 (1987) 3607; G. Auriemma, M. Felcini, P. Lipari and 
J. L. Stone, Phys. Rev. D 37 (1988) 665; A. Nicolaidis, Phys. Lett. B 200 
(1988) 553; J. M. LoSecco, Phys. Rev. D 47 (1993) 2032; J. M. Gelb, W.-K. Kwong 
and S. P. Rosen, ref. \cite{GKR}; Q. Y. Liu, M. Maris and S. T. Petcov, Phys. 
Rev. D 56 (1997) 5991; M. Maris and S. T. Petcov, Phys. Rev. D 56 (1997) 7444; 
M. Maris and S. T. Petcov, Phys. Rev. D58 (1998) 113008. 



\end{thebibliography}
\end{document}